# PKI Implementation Issues:
# A Comparative Study of Pakistan with some Asian Countries

**Nasir Mahmood Malik, Tehmina Khalil, Samina Khalid, Faisal Munir Malik**
Department of Computer Science, Bahria University, Islamabad, Pakistan
emailnasir@yahoo.com, tehmina_khalil08@yahoo.com, noshi_mir@yahoo.com, faisy20@yahoo.com

*Abstract* -The paper includes Public Key Infrastructure (PKI), its need and requirements and introduction of some renowned PKI products. However, the major thrust of this work is that how PKI can enhance security of various systems. The paper is intended to serve as a guide on how to adequately prepare for some of the challenges that may be encountered especially in developing countries like Pakistan. The detail of PKI implementation issues is also included in the paper along with future challenges regarding implementation of PKI. Furthermore, paper includes technical issues hindering the implementation of PKI through comparison of PKI issues in Pakistan and some of Asian countries mainly Taiwan, Japan and Singapore. The paper also highlights the PKI issues and learnt lessons regarding PKI implementation and can act as a comprehensive guide for successful future PKI deployments.

*Keywords: Public Key Infrastructure, PKI, Issues, Cryptography, Survey, Survey Comparison*

## 1. Introduction

In contrast to physical world, internet is anonymous and it is very difficult to find out who is at the other end of the communication. PKI is a widely accepted global standard for Internet security today, as the major challenges for online communication include establishing an online trust similar to physical marketplace and communication binding contracts for online transactions. That is why information security is a wide area of interest for researchers today but the implementation of PKI is a challenge for most of the developing countries. Also to achieve interoperability it is a necessity that countries should establish PKI as per the widely accepted standards. This paper is an attempt to provide guidelines for PKI implementation and to investigate various issues in Pakistan and comparison of this analysis with some of the similar case studies of Asian countries mainly Taiwan, Japan and Singapore. Organization of the paper is as follows, the second section introduces cryptography, PKC and PKI with some of its products. In the third section, analysis of PKI status in Pakistan is explained. Fourth and fifth sections give background information of this survey, methodology and comparison with some of the Asian countries. Finally, conclusions drawn, few recommendations and future actions are given.

## 2. Cryptography

Cryptography is mostly associated with encryption/decryption and there are two main categories of cryptography i.e. symmetric cryptography using a secret-key and asymmetric cryptography using a public key based encryption algorithms respectively. The difference between these algorithms is that symmetric algorithms use the same key for encryption and decryption (or that the decryption key is easily derived from the encryption key), whereas asymmetric algorithms use a different key for encryption and decryption, and the decryption key can not be derived from the encryption key [1]. Typical asymmetric algorithms compared in the forthcoming sections include DSA (Digital Signature Algorithm), Diffie-Hellman (DH), and RSA (Ron Rivest, Adi Shamir and Len Adleman). By using the characteristics of these symmetric and asymmetric algorithms various PKI products are developed by different vendors which are widely available in the market.

### 2.1 Public Key Cryptography (PKC)

Public Key Cryptography, also known as asymmetric cryptography, is a form of cryptography in which the key used to encrypt a message differs from the key used to decrypt it. In PKC, a user has a pair of public key and private key. The private key is kept secret, while the public key may be widely distributed. Incoming messages are encrypted with the recipient's public key and can only be decrypted with his/her corresponding key i.e. private key. The keys are related mathematically, but private key cannot be practically derived from the public key, mostly, asymmetric algorithms are used in PKC [2].

### 2.2. Public Key Infrastructure (PKI)

A framework for creating a secure method for exchanging information based on public key cryptography. The foundation of a PKI is the certificate authority (CA), which issues digital certificates that authenticate the identity of organizations and individuals over a public system such as the Internet. The certificates are also used to sign messages, which ensure that messages have not been tampered with [3]. Due to these services provided by PKI many PKI products are developed and some of the renowned products are given in the forthcoming section.

### 2.3. PKI Products

There are different ways to get PKI services, since a variety of PKI products and vendors are available [4], [5], [6]. But the fact is that we don't have any perfect and generic solution that could address all the ten issues of PKI given in [7]. However, we can differentiate these products on the basis of some technological differences like how certificates are issued, agility [8], deployed, maintained and revoked. Some of popular PKI products include Entrust, VeriSign and RSA Security.

### 3. PKI Status in Pakistan

Traditional methods of signing agreement orders, etc. must be reproduced electronically as well. PKI provides the means to do this. Proper handlings of legal implications involved in electronic transactions are most important part of PKI implementation. Binding customers and businesses to contracts is the act of non-repudiation. PKI should be deployed based on some law e.g. digital signature should be generated according to some digital signature law and then somebody must be made liable if something goes wrong. Also owner of a public key certificate can not repudiate a signature that is generated with the appropriate signing key. The major components regarding the current legal status of PKI implementation in Pakistan include:

Electronic Transaction Ordinance (ETO) 2002: To recognize digital documents, certificates and signatures as equivalent to paper documents and written signature but this Ordinance has not been updated regularly. The legislation recognizes all files/data in any electronic format as documents and that these documents shall not be denied legal power and enforceability.

Electronic Crimes Act 2004: This Act provides laws for punishment of the electronic crimes and for accompanying matters. The major issue is the regular update of this Act as cyber crime techniques keep on getting sophisticated. Electronic Certification Accreditation Council (ECAC): ECAC provides conducive legal and policy framework that creates an environment of trust, predictability and certainty in the country. CA's working in Pakistan: One of the major CA working in Pakistan is National Institutional Facilitation Technologies (NIFT) formed in early 1995 is working as partner with VeriSign and share VeriSign CA services. PK-GRID-CA, owned by Quaid-i-Azam University, issues X.509 digital certificates to use grid resources under secure environment. It is serving around 10,000 Pakistani scientists all around the world.

### 4. Survey Background

We surveyed more than 133 local organizations having national as well as international business. Also, major focus was on IT & Telecom/ related companies But in this paper we have only included those organizations were PKI implementation exists upto some level i.e. 55. As shown in Figure-1, 43%, 26%, 17% and 14% targeted organizations belong to IT, Financial/Banking, Government/Public and Telecom sectors respectively.

**Figure-1: Target Organizations**

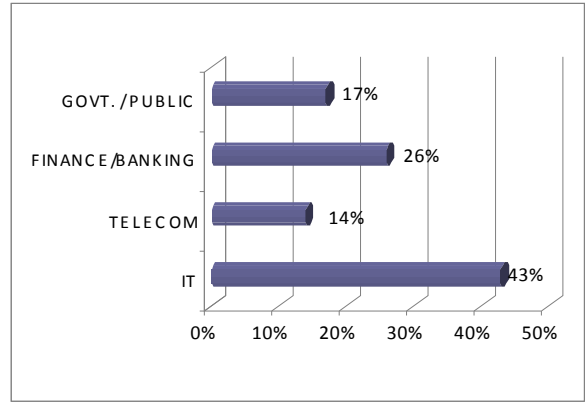

But out of these 55 organizations as explained in Figure-2, 20 (36%) organizations have not yet implemented PKI but do plan to implement in the near future whereas 35 organizations (64%) were at various stage of PKI implementation. Among those 35 organizations, PKI implementation was in progress in 20 (57%) while 15 (43%) organizations have already completed PKI implementation.

**Figure-2: PKI Survey Background**

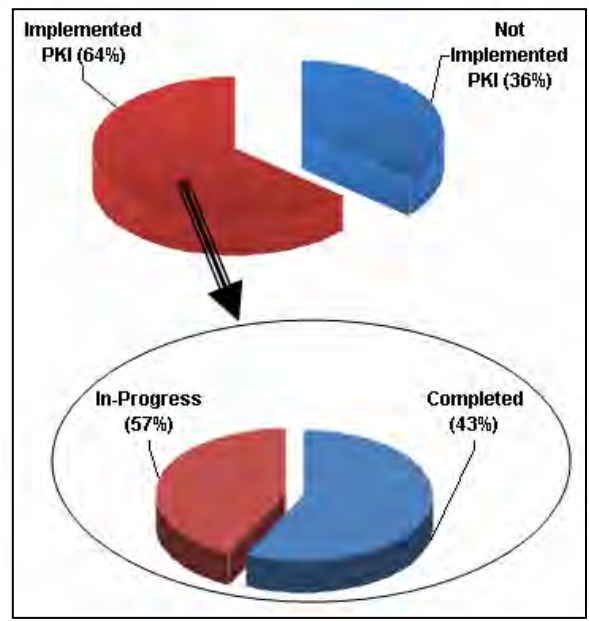

Figure-3 shows that the percentage of organizations which responded that they have developed PKI application on their own is much lower (25%) than the percentage of those which replied that they have not developed PKI application on their own (75%). This shows that most of the organizations are comfortable when some third party is involved in this process. This is one of the major reasons that most number of IT sector companies have already deployed PKI in their respective organizations and are providing these solutions to its customers as well.

**Figure-3: PKI Applications**

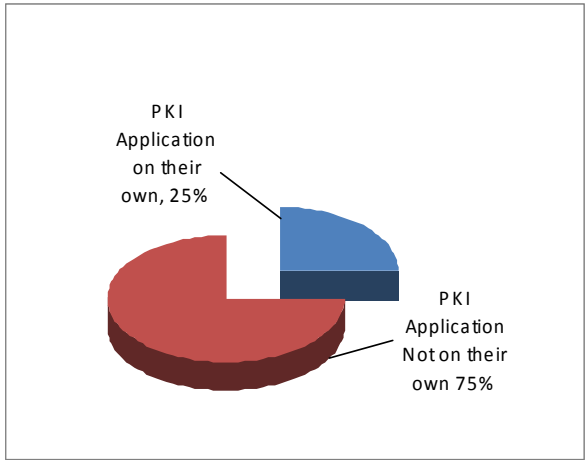

Only those respondents who have completed PKI implementation in their organizations were further asked to answer their PKI deployment and related information regarding it. The detail of their responses has been compiled and analysis of this comparison is given in Table-1.

## 5. Comparative Analysis of Issues

PKI has become the object of international attention and much has been done to realize national and international standards for PKI, for example X.509. There are, however, serious PKI implementation issues as different countries and different organizations may adopt different security policies, implementations and standards. This raises the question of interoperability between these various implementations, especially in such a way as to create a global trust domain [9]. Asia PKI Forum is an international organization, set up in June 2001 with the aim of establishing PKI interoperability in Asia. There are seven countries, China, Japan, Korea, Singapore, Hong Kong, Macao, and Chinese Taipei that are members of this Asia PKI Forum [10]. Out of these member countries Japan, Singapore and Chinese Taipei PKI Forum has conducted PKI surveys and our questionnaire is mostly based on these surveys [11], [12], [13]. We analyzed the questionnaire responses in term of the following aspects implementation status of PKI, applications, products, challenges and obstacles faced during implementation of PKI in Pakistan. As per our survey results, Taiwan has an edge when it comes to PKI implementations as 65% of organizations of Taiwan are having PKI where as in Singapore, Japan, and Pakistan 56%, 42% and 25% organizations have implemented PKI respectively. In Pakistan PKI implementations are at initial stages due to lack of technical knowledge. The main reason for higher PKI deployment in Taiwan is due to compliance of legal and business requirements. This is clearly manifest as the requirements enforced by the business partners and legal requirements in Taiwan are 70% and 29% with higher percentage as compared to Japan, Singapore and Pakistan. This shows that if an organization wants to survive in Taiwan it is more likely to deploy PKI as compared to an organization doing business in Japan, Singapore or in Pakistan. Moreover, the PKI deployment is mainly done in Pakistan to fulfill security needs as 52% organizations implemented PKI to meet security requirements. Also in Taiwan and Japan PKI is deployed to meet security requirements as 92% organizations in both countries are using PKI to meet security needs.

### Table 1: COMPARISON OF PKI IMPLEMENTATION IN PAKISTAN

| DISCRIPTION | PAKISTAN | SINGAPORE | JAPAN | TAIWAN |
|---|---|---|---|---|
| Organizations Implemented/not Implemented PKI | | | | |
| Organizations having PKI | 25% | 54% | 42% | 65% |
| Organizations not having PKI | 75% | 46% | 54% | 24% |
| **PKI Based Application** | | | | |
| Cross Authentication-SSL | 35% | 26% | 69% | 73% |
| Web Application Software | 15% | 41% | 38% | 76% |
| VPN | 10% | 18% | 39% | 27% |
| Secure E-mail | 40% | 15% | 31% | 51% |
| **PKI Functionalities Utilized** | | | | |
| Authentication | 17% | 29% | 100% | 89% |
| Data Integrity | 30% | 32% | 54% | 84% |
| Confidentiality | 18% | 20% | 84% | 84% |
| Non-Repudiation | 35% | 19% | 38% | 92% |
| **PKI Brands** | | | | |
| VeriSign | 25% | 60% | 19% | 14% |
| RSA | 31% | 12% | 8% | 27% |
| **Reasons for PKI Deployment** | | | | |
| Security | 52% | 9% | 92% | 92% |
| Business Partner's Demand | 20% | 19% | 31% | 70% |
| Legal Requirement | 16% | 22% | 15% | 29% |
| **Obstacles in Implementation of PKI** | | | | |
| Lack of Technical Knowledge | 24% | 30% | 46% | 35% |
| Limited Options of PKI Products | 18% | 25% | 23% | 38% |
| Integration Difficulty | 16% | 19% | 15% | 30% |
| **Key Utilization** | | | | |
| Single Key | 33% | 77% | 53% | 8% |
| Dual Key | 67% | 23% | 8% | 67% |
| **PKI Implementation Duration** | | | | |
| 1 Year OR Less | 7% | 34% | 36% | 40% |
| 1 to 2 Years | 40% | 25% | 7% | 3% |

The major PKI protocol implemented in Taiwan and Singapore is in web application software 76% and 41% respectively. Also, web applications are more secure in Taiwan and Singapore as compared to Japan and Pakistan. Cross-Authentication-SSL protocol is almost equally used in Japan and Taiwan i.e. 69% and 73% respectively. PKI

Protocol mostly used by Pakistani organizations is secure e-mail i.e. 40% that indicates that in Pakistan PKI is primarily used for secure e-mail communication. As far as utilization of PKI functionalities is concerned, organizations in Japan are putting more emphasis on authentication (100%), whereas, in Taiwan and Pakistan the main stress is on non-repudiation 92% and 35% respectively. While in Singapore, Data Integrity function of PKI is mostly utilized (32%). When it comes to PKI brands/vendors, Taiwanese and Pakistani organizations prefer RSA (27% and 31% respectively), whereas in Japan and Singapore VeriSign is mostly used i.e. 19% and 60% respectively. The major obstacles in Taiwan are only the limited options of PKI products and solutions (38%).

Furthermore, Taiwanese organizations face more problems while integrating the PKI implementation with the existing IT infrastructure i.e. 30% as compared to only 15%, 16% and 19% in Japan, Pakistan and Singapore respectively. This shows that organizations in Taiwan are deploying PKI solutions due to legal requirements instead of the difficulties of integration. Also, 65% of Taiwanese organizations have already implemented PKI in contrast to only 42% of Japan's organizations. But, the major plus with Taiwan is that more technical knowledge is available and only 35% organizations say that they have not implemented PKI due to lack of technical knowledge in contrast to 46% of Japanese organizations admit the lack of technical knowledge. PKI implementation cycle is much shorter in Taiwan. About 40% of the Taiwanese organizations say that the cycle is completed in less than a year as compared to Japan and Singapore (36% and 34%). While in Pakistan PKI implementation cycle takes much more time as 40% of Pakistani organizations complete PKI implementation in 1 to 2 years.

Lastly, the predominant use of dual key utilization in PKI implementation in Taiwan and Pakistan is linked with the security requirements and business partner requirements as 67% organizations in both the countries are using dual key, as compared to only 8% and 23% in Japan and Singapore respectively. Dual Key utilization is safer and secure way of PKI implementation, while 53% and 77% organizations in Japan and Singapore are using single key pair. We can safely say that implementation of PKI is more mature, advanced and safe in Taiwan rather than in Japan, Singapore and Pakistan. The vast majority thinks that Taiwan will be the country with most potential for mutual recognition in PKI technology.

## 6. Conclusion

PKI Survey of Pakistan has shown that this technology has not reached to a wider number of organizations/users. The major reason is that there is no commercialy available Certificatin Authority (CA) in Pakistan. As per our survey results, the main reason that Pakistani organizations are not going towards PKI implementation is because of it's high cost. PKI is considered important for cross border trade in Pakistan as 53% of organizations are using PKI for communicating with foreign trading partners. Also, most of the organizations are planning to implement PKI in near future. Furthermore, lack of technical knowledge in this field has also made it much harder for PKI projects to be considered as viable by the organizations as it does take some basic technological know how to use it. In Pakistan IT industry, Banks and Financial Institutes make up most of the percentage that deploy PKI. While in other sectors security implementation is not a priority as they do not consider it a necessity. Finally, Cross Authenticaton-SSL and Secure E-mail are the most recognized potential applications of PKI in Pakistan especially for trading with foriegn partners. Moreover, cyber security laws need to be updated on regular basis. But countries like Pakistan do face lots of issues in implementation of these laws throughout the country especially e-Laws e-Transactions, e-Crimes, digital signatures, digital certicate, digital forensics etc are at very early stage of implementation. Lastly, digital signature law is still not enforced in the country and acceptance of digital documents is not a norm in Pakistan. However, the situation is improving and couple of public and private sector organizations have implemented "paper-less" initiatives.

## Future Work

As per our research, Pakistan has to go a long way to achieve sufficent cyber security solutions. So, there is an opportunity to explore other security (other than PKI) options. Also, the major issue for emerging Pakistani market is the establishment of commercial CA as the country can not rely on costly PKI products from outside Pakistan for even along with regular updation of cyber laws. We are already working on non-PKI cyber security solutions and feasiblity study of establishment of commercial CA in Pakistan.